# On the appearance of a Dirac delta term at the origin in the Schrödinger radial equation

J. Etxebarria

*Department of Condensed Matter Physics, University of the Basque Country*
*UPV/EHU, 48080 Bilbao, Spain*

E-mail: j.etxeba@ehu.es

## Abstract

We revisit a recent discussion about the boundary condition at the origin in the Schrödinger radial equation for central potentials. Using a slight modification of the usual spherical coordinates, the origin of a previously reported Dirac delta function term at the origin is clearly shown up. As a consequence, a vanishing boundary condition at the origin must be imposed in solving the radial equation, regardless the kind of potential.

## 1. Introduction

The stationary states of central potentials are usually represented by a product of an angular part (spherical harmonics) and a radial part $R$. The latter is calculated as a solution of the radial equation

$$-\frac{\hbar^2}{2m}\frac{d^2 y}{dr^2} + \left[V(r) + \frac{l(l+1)\hbar^2}{2mr^2}\right]y(r) = Ey(r) \qquad (1)$$

where $y(r)=R(r)r$ $(r>0)$, $E$ is the energy, $m$ the mass of the particle and $l$ the azimuthal quantum number. Eq. (1) is solved with the supplementary condition

$$y(0)=0. \qquad (2)$$

Though the boundary condition (2) is used in virtually all textbooks on quantum mechanics it has not always been justified correctly. In fact this question is discussed differently in different textbooks. For example Schiff [1] (page 84), and Landau [2] (page 104), use the argument that the wave function must be finite at the origin, whereas Powell and Crasemann [3] (page 222), suggest that it is the normalization of the wave function that leads to (2). The normalization however can be realized even if the wave function is singular at the origin, i.e., if y(0) ≠0. Griffiths [4] (page 141), recognizes the subtlety of this issue and avoids a direct consideration of the problem. Messiah [5] (page 346), defines the Hermitian radial momentum and its Hermitian property requires this boundary condition. Another explanation is that given by Flügge [6] (page 157). He indicates that y(0) ≠0 would lead to a divergent energy integral. All these examples point out that the rational for (2) is not often clear.

Recently, Khelashvili and Nadareishvili [7] have given a succinct and convincing reason for (2), making explicit an argument already outlined by Dirac in his classical book [8] (page 156). These authors have shown the appearance of a Dirac delta function term in the radial Schrodinger equation when the Laplace operator is written in spherical coordinates,

$$-\frac{\hbar^2}{2mr}\frac{d^2y}{dr^2} + \frac{2\pi\hbar^2}{m}y(r)\delta(\mathbf{r}) + \left[V(r) + \frac{l(l+1)\hbar^2}{2mr^2}\right]\frac{y(r)}{r} = E\frac{y(r)}{r} \qquad (3)$$

This additional term had been previously unnoticed in the literature, and gives a clear explanation for the necessity of imposing the supplementary condition (2) for solving the usual radial equation (1).

This conclusion is achieved regardless of whether or not the nature of the potential is regular or singular at the origin [if close to $r=0$ $V(r) \propto 1/r^\alpha$, the potential is called regular (singular) when $\alpha < 2$ ($\alpha \geq 2$)]. In this note we will show a very simple justification of (3). The argument can be easily understood for undergraduate students of physics.

## 2. Appearance of the Dirac delta term

For our purposes we will use slightly-modified spherical coordinates. Usually $r$ ranges from 0 to ∞, and the polar angle $\theta$ ranges from 0 to π. However, in order to analyze the problems at the origin it is convenient to take a new coordinate $r$ that ranges from -∞ to +∞, and a new $\theta$ varying from 0 to π/2. This choice unambiguously describes the position of an arbitrary point. Simply, the position vectors of the Northern hemisphere have now coordinates $r>0$, and those of the Southern hemisphere $r<0$. This choice has been shown to be useful to avoid "losing" δ functions on calculating fields of point charges and dipoles [9]. To prevent confusion the usual radial coordinate will be denoted by |r|.

The relationship between the usual spherical coordinates $(|r|,\theta_u,\varphi)$, and the modified coordinates $(r,\theta,\varphi)$ is

$$r = \begin{cases} |r| \text{ if } \theta_u < \pi/2 \\ -|r| \text{ if } \theta_u > \pi/2 \end{cases}$$

$$\theta = \begin{cases} \theta_u \text{ if } \theta_u < \pi/2 \\ \pi - \theta_u \text{ if } \theta_u > \pi/2 \end{cases}.$$

We first consider a wave function without angular dependence (*l*=0). The essence of the argument already appears in this simple case. The Laplace operator can be written as [9]

$$\nabla^2 = \frac{1}{r}\frac{d^2}{dr^2}r \qquad (4)$$

A wave function without angular dependence (or any function with only radial dependence) is described by an *even* function of the new coordinate *r*, i.e., *R(r)=R(-r)(=R(|r|))*, see Fig. 1a. The central potential is also given by an *even* function *V(r)=V(-r)(=V((|r|)))*. In contrast, *y(r)= rR(r)* is *odd*.

Now, when computing $\nabla^2 R$ using (4) it is convenient to notice the following points in the different steps of the calculation:

i) *rR(r)* is odd, and may have a discontinuity (Fig. 1b) if its limits on the right and left hand sides around *r*=0 are different, i.e., if $y(0^+) \neq y(0^-)(=-y(0^+))$

ii) *d(rR)/dr* is even, and may have a δ- type discontinuity at the origin, whose strength is given by the magnitude of the discontinuity of *rR* at the origin. In other words, there appears a $a\delta(r)$ term, with $a = 2y(0^+)$. This is the key point of the argument.

iii) $d^2(rR)/dr^2$ is odd, and has a contribution $a[d\delta(r)/dr] = -a\delta(r)/r$ at the origin [10].

iv) Finally, $\nabla^2 R$ is even and contains an extra term $-2y(0^+)\delta(r)/r^2$.

By calculating $\nabla^2 R$ for *r*>0 in the conventional way, the Schrodinger equation,

$$\left[-\frac{\hbar^2}{2m}\nabla^2 + V(r)\right]R(r) = ER(r) \qquad (5)$$

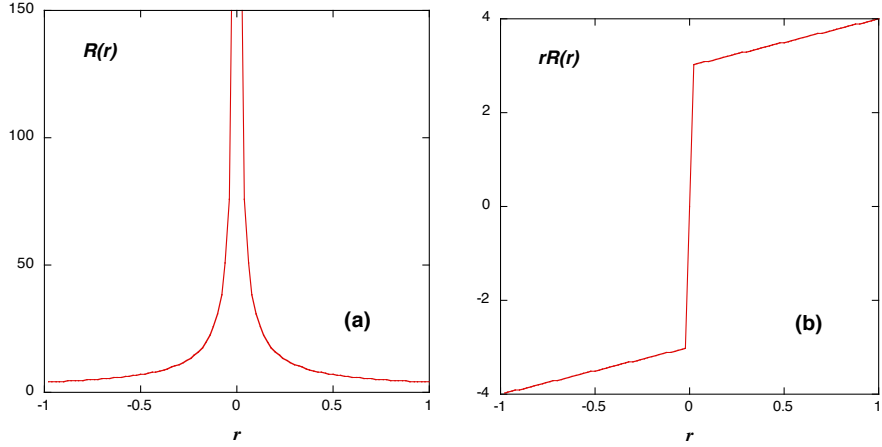

**Figure 1**.- Radial dependence of a hypothetical wave function without angular dependence $R(r)$ (a), and $rR(r)$ (b) showing a discontinuity at $r=0$. The derivative of $rR$ has a $\delta$ term in the origin. All scales are in arbitrary units.

gives rise to

$$-\frac{\hbar^2}{2mr}\frac{d^2 y}{dr^2} + \frac{\hbar^2}{mr^2}y(0^+)\delta(r) + V(r)\frac{y(r)}{r} = E\frac{y(r)}{r} \ . \qquad (6)$$

The $\delta(r)$ term can be expressed using a three-dimensional delta $\delta(\mathbf{r})$ since, when the latter is multiplied by a function with only radial dependence, its effect is the same than the multiplication by $\frac{\delta(r)}{2\pi r^2}$. In other words, we have

$$\delta(\mathbf{r}) = \frac{\delta(r)}{2\pi r^2} . \qquad (7)$$

This identity is shown by explicit integration of the product of $\delta(\mathbf{r})$ with a test function $f(r)$

$$\iiint_{\text{all space}} \delta(\mathbf{r})f(r)d^3\mathbf{r} = \int_0^{2\pi}d\varphi\int_0^{\pi/2}d\theta\sin\theta\int_{-\infty}^{+\infty}\delta(\mathbf{r})f(r)r^2 dr = \int_0^{2\pi}d\varphi\int_0^{\pi/2}d\theta\sin\theta\int_{-\infty}^{+\infty}\frac{\delta(r)}{2\pi r^2}f(r)r^2 dr = f(0)$$

Finally, (6) and (7) lead us to

$$-\frac{\hbar^2}{2mr}\frac{d^2y}{dr^2}+\frac{2\pi\hbar^2}{m}y(r)\delta(\mathbf{r})+V(r)\frac{y(r)}{r}=E\frac{y(r)}{r}. \tag{8}$$

Eq. (3) is obtained from (8) in the case $l \neq 0$ by adding, in the usual way, the centrifugal term to the actual potential. This point is further justified in the Appendix.

## 3. Conclusion

Thus, unless $y(0)=0$, the radial equation (3), which rigorously follows from the Schrödinger equation, is not exactly coincident with the usual radial equation (1). Therefore, condition (2) is compelled for a consistent use of Eq. (1). If we take $y(0) \neq 0$ to solve (1), we are implicitly dealing with a problem of a potential to which an extra term $\frac{2\pi\hbar^2}{m}r\delta(\mathbf{r})$ is artificially added. The only way to avoid the delta term is a vanishing boundary condition at the origin. This conclusion is valid regardless of whether the nature of the potential is regular or singular.

**Appendix: The case $l \neq 0$**

If $l \neq 0$ we will see that the solutions of the whole Schrödinger equation, $R(|r|)Y_l^m(\theta_u,\varphi)$, are written as

$$R(r)Y_l^m(\theta,\varphi) \tag{A1}$$

in the new coordinate system. This is evident for $\theta_u < \pi/2$. On the other hand, the values of $R(r)$ for negative $r$ ($\theta_u > \pi/2$) can be obtained by using the parity property of the spherical harmonics as follows:

Upon an inversion, a point of coordinates $(r,\theta,\varphi)$ is transformed into $(-r,\theta,\varphi+\pi)$. Thus we must have $R(-r)Y_l^m(\theta,\varphi+\pi)=(-1)^l R(r)Y_l^m(\theta,\varphi)$, because the spherical harmonic has parity $l$. Since, on the other hand,

$Y_l^m(\theta,\varphi+\pi)=(-1)^m Y_l^m(\theta,\varphi)$, we conclude that $R(r)$ must have a definite parity, i.e.,

$$R(-r) = (-1)^{l+m} R(r)). \tag{A2}$$

In other words, the stationary states are of the form (A1) with the condition (A2).

On the other hand, the Laplace operator has the form

$$\nabla^2 = \frac{1}{r}\frac{\partial^2}{\partial r^2} r - \frac{L^2}{\hbar^2 r^2}$$

where $L^2$ is the angular momentum operator. Then the full Schrodinger equation

$$\left[-\frac{\hbar^2}{2m}\nabla^2 + V(r)\right]\left[R(r)Y_l^m(\theta,\varphi)\right] = ER(r)Y_l^m(\theta,\varphi)$$

leads to

$$-\frac{\hbar^2}{2m}\frac{1}{r}\frac{d^2}{dr^2}(rR(r)) + \left(V(|r|) + \frac{l(l+1)\hbar^2}{2mr^2}\right)R(r) = ER(r)$$

both for positive and negative $r$. Finally, using (A2) we arrive at

$$-\frac{\hbar^2}{2m}\frac{1}{r}\frac{d^2}{dr^2}(rR(|r|)) + \left(V(|r|) + \frac{l(l+1)\hbar^2}{2mr^2}\right)R(|r|) = ER(|r|)$$

which is like Eq. (5) but with the additional centrifugal term. Following the same steps given in the main text we deduce the analogue to Eq. (6),

$$-\frac{\hbar^2}{2mr}\frac{d^2 y}{dr^2} + \frac{\hbar^2}{mr^2}y(0)\delta(r) + \left[V(r) + \frac{l(l+1)\hbar^2}{2mr^2}\right]\frac{y(r)}{r} = E\frac{y(r)}{r}$$

which, in its turn, gives rise to (3).